\documentclass[11pt, a4paper,  notitlepage]{extarticle}

\usepackage[left=3.5cm,right=3cm, top=3cm,bottom=3cm,bindingoffset=0cm]{geometry}
\usepackage{graphicx}
\usepackage{graphicx}
\usepackage{mathtools}
\usepackage[affil-it]{authblk}
\begin{document}

\title{Bitcoin Price Predictive Modeling \\ Using Expert Correction}
\author{Bohdan M.  Pavlyshenko \\  \small{Ivan Franko National University of Lviv, Ukraine \\ b.pavlyshenko@gmail.com,  www.linkedin.com/in/bpavlyshenko/}}
\maketitle

\begin{abstract}
The paper studies the linear model for Bitcoin price which includes regression features based on Bitcoin currency statistics, 
mining processes, Google search trends, Wikipedia pages visits. 
The pattern  of deviation of regression model prediction  from real prices is simpler comparing to price time series. It is  assumed
that this pattern can be predicted by an experienced expert. In such a way, using the combination of the regression model and 
expert correction, one can receive better results than with either regression model or expert opinion only. 
It is shown that Bayesian approach makes it possible to  utilize the probabilistic approach using distributions with fat tails and take into account the outliers in Bitcoin 
price time series.
\end{abstract}

\section{Introduction}
One of the main goals in the Bitcoin analytics is price forecasting. There are many factors which influence the price dynamics. 
The most important factors are: interaction between supply and demand, attractiveness for investors, financial and macroeconomic indicators, technical indicators such as difficulty, 
the number of  blocks created recently, etc. A very important impact on the cryptocurrency price has trends in social networks and search engines. 
Using these factors, one can create a regression model with good fitting of bitcoin price on the historical data. 
Paper \cite{kristoufek2013bitcoin} shows that views of Wikipedia Bitcoin related pages correlate with Bitcoin price movements. It reflects
the potential investors' interests in  cryptocurrency. Google trends of search Bitcoin related keywords show different effects, including investors' interests, 
speculators' activities, etc.   
In ~\cite{bouoiyour2015bitcoin}, Bitcoin price was analyzed.
The paper ~\cite{bouoiyour2016drives} studies different drivers of Bitcoin price.
In ~\cite{matta2015bitcoin}, a significant  correlation  between Bitcoin price and social and web search media trends was shown.
In ~\cite{dyhrberg2016bitcoin},  the analysis shows that bitcoin has many similarities to both gold and the dollar. 
In ~\cite{shah2014bayesian},  the use of Bayesian regression for Bitcoin price analytics was  studied.  
The impact of news on investors' behavior  is studied in ~\cite{barber2007all}. 
Bitcoin economy, behavior and mining are considered in ~\cite{grinberg2012bitcoin, kroll2013economics}. 
Bitcoin market behaviour, especially  price dynamics is the subject of different studies ( \cite{ciaian2016economics, kristoufek2013bitcoin}).  
Different factors affecting bitcoin price are analyzed in the \cite{ciaian2016economics}. 
The specific feature of Bitcoin is that this cryptocurrency is neither issued nor controlled by financial or political institutions such as Central Bank, government, etc. 
 Bitcoin is being mined without economic underlying factors.  
In ~\cite{ciaian2016economics}, the economics of BitCoin Price Formation has been analyzed.   
Price dynamics is mostly affected by speculative behaviour  of investors. One of the main bitcoin drivers is the news in the Internet. 
According to efficient market theory, the stock and financial markets are not predictable, since all available information 
is already reflected in stock prices. But nowdays the dominance of efficient market theory is not so obvious.  Some influential scientists
argue that market can be partially predictable  (~\cite{malkiel2003efficient}). Behind modern market prediction, there are behavioral and psychological theories.
Some economists believe that historical prices, news, social network activities contain patterns that make it possible to partially predict financial market. 
Such theories and approaches are considered in the survey ~\cite{malkiel2003efficient}. 

In this paper, we consider an approach for building regression predictive model for bitcoin price using expert correction by adding a correction term.
It is assumed that an experienced expert can make model correction relying on his or her experience. 
\section{Bitcoin Price Modeling}
\begin{figure}
\centerline{\includegraphics[width=1\textwidth]{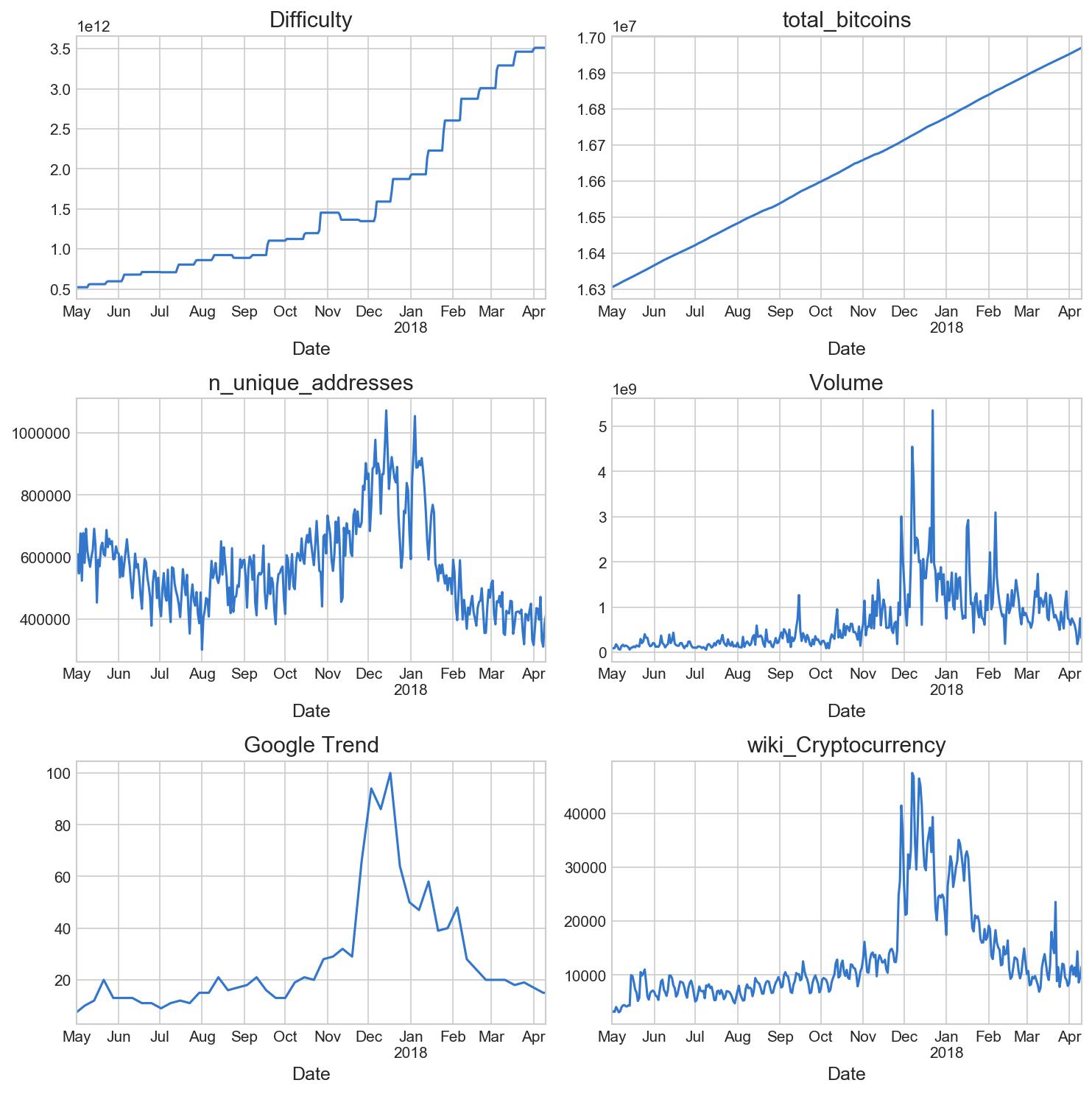}}
\caption{Time series of features}
\label{fig1}
\end{figure}

\begin{figure}
\centerline{\includegraphics[width=1\textwidth]{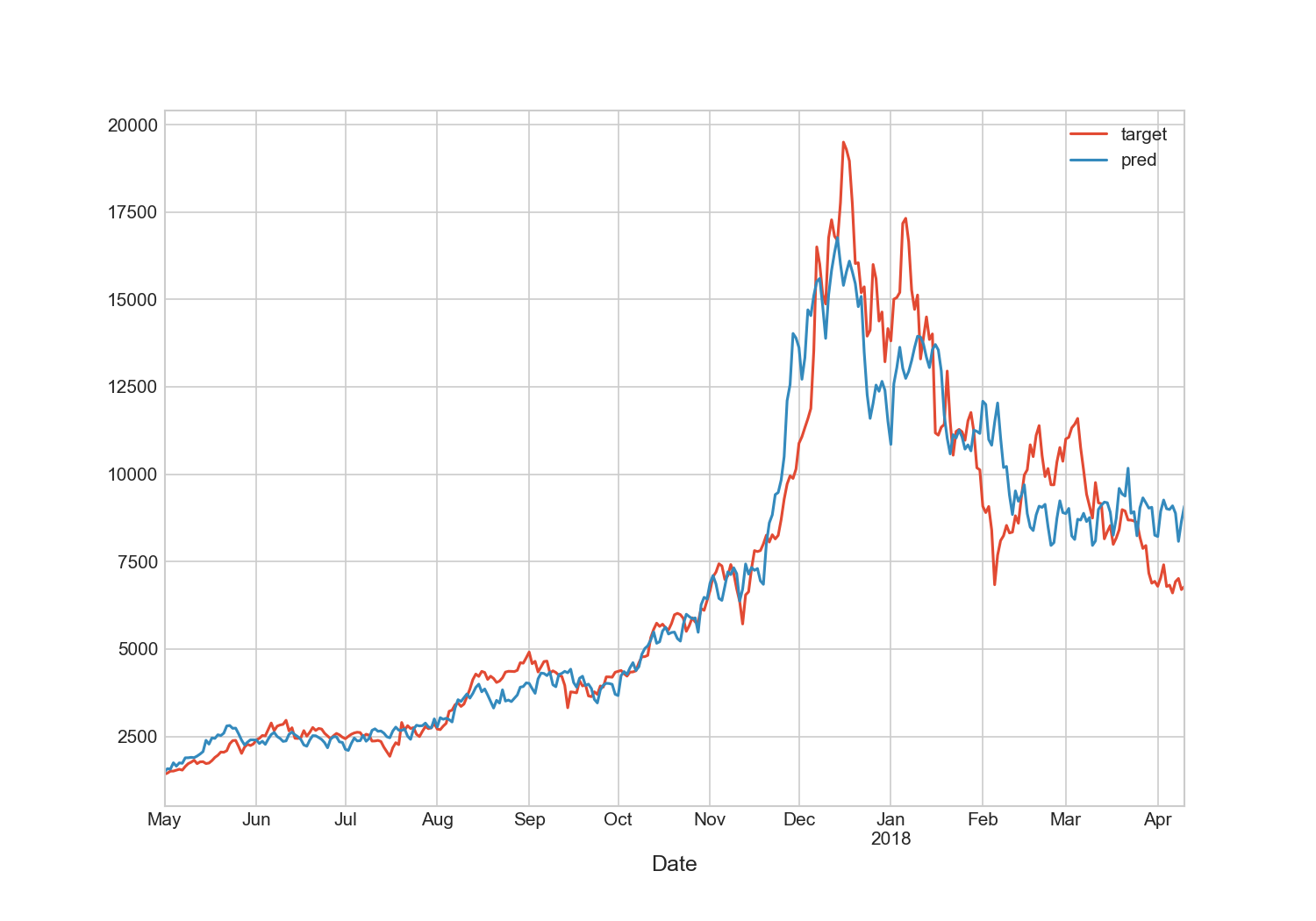} }
\caption{Bitcoin price dynamics and prediction}
\label{pred1}
\end{figure}

\begin{figure}
\centerline{\includegraphics[width=0.75\textwidth]{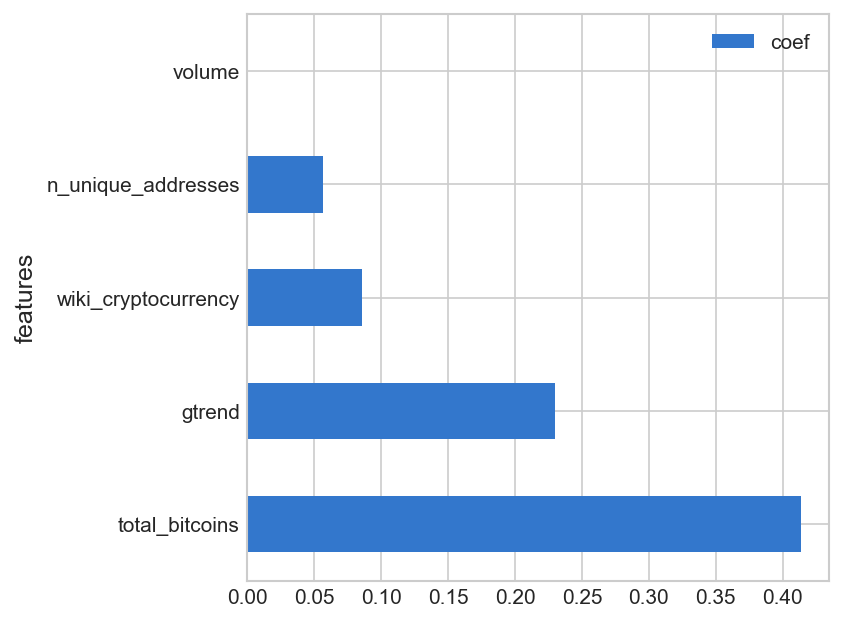}}
\caption{Features regression coefficients}
\label{coef1}
\end{figure}

Let us consider a regression model for Bitcoin price. 
As the regressors in our model, we used historical data which describe Bitcoin currency statistics, 
mining processes, Google search trends, Wikipedia pages visits. 
As the currency statistics,  we took 
$total\_bitcoins$ - the total number of bitcoins that have already been mined,
$price$ - average USD market price across major bitcoin exchanges,
$volume$ -the total USD value of trading volume on major bitcoin exchanges.
As mining information, we took 
$difficulty$, which is a relative measure of how difficult it is to find a new block.
As network activity, we took $n\_unique\_addresses$  which is the total number of unique addresses used on the Bitcoin blockchain.
Time series for mentioned above variables was taken from Bitcoin.info site. 
As the factors of price formation, we also considered Google trend for 'bitcoin'  keyword and  number of visits of  Wikipedia page about 'cryptocurrency'.
Time series for chosen features are shown on the Fig.~\ref{fig1} .
A target variable $price$ and all the regressors are considered in the logarithmic scale, so  $price'=ln(price+1)$, $x_{i}'=ln(x_{i}+1)$ 
After logarithmic transformation all the regressors were normalized by extracting mean values and then dividing by standard deviation. 
We can write the regression model  as:
\begin{equation}
\label{e1}
\begin{split}
&price'=\alpha+\beta_{gtrend}\cdot gtrend'+ \\
&\beta_{wiki\_cryptocurrency}\cdot wiki\_cryptocurrency'+ \\
&\beta_{difficulty}\cdot difficulty + \\
&\beta_{n\_unique\_addresses}\cdot n\_unique\_addresses' + \\
&\beta_{total\_bitcoins}\cdot total\_bitcoins' +\\
& \beta_{volume}\cdot volume' 
\end{split}
\end{equation}
For the numerical modeling, we used Python with the packages \textit{pandas, numpy, scipy, matplotlib, seaborn , sklearn, quandl, pystan} . 
To get time series of Wikipedia pages visits, we used Python package \textit{mwviews}.
To find coefficients $\alpha, \beta_i$, we used linear regression with Lasso regularization from python package scikit-learn. 
Fig.~\ref{pred1} shows real dynamics of Bitcoin price  and the price predicted by  regression model (~\ref{e1}). 
Fig.~\ref{coef1} shows obtained linear regression coefficients for features. 
The results show the importance of Google trends for Bitcoin keyword searches, and the importance of views of Wikipedia pages about the cryptocurrency. 
For this analyzed model,  we received the error value RMSE=1277.8. 

\section{Modeling of expert correction}
\begin{figure}
\centerline{\includegraphics[width=0.75\textwidth]{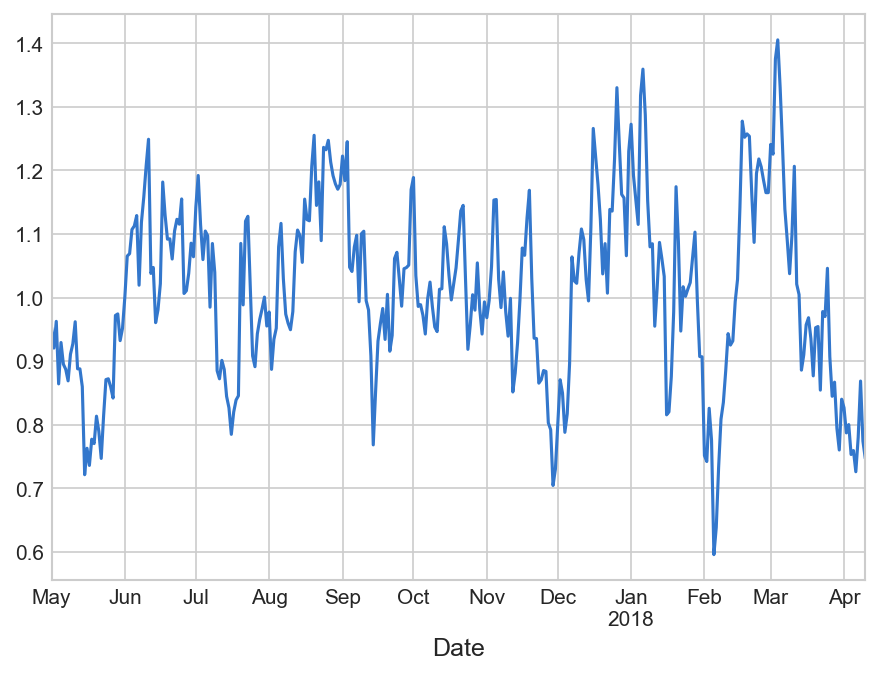}}
\caption{The ratio of real Bitcoin price to predicted price}
\label{targetpred}
\end{figure}

\begin{figure}
\centerline{\includegraphics[width=0.75\textwidth]{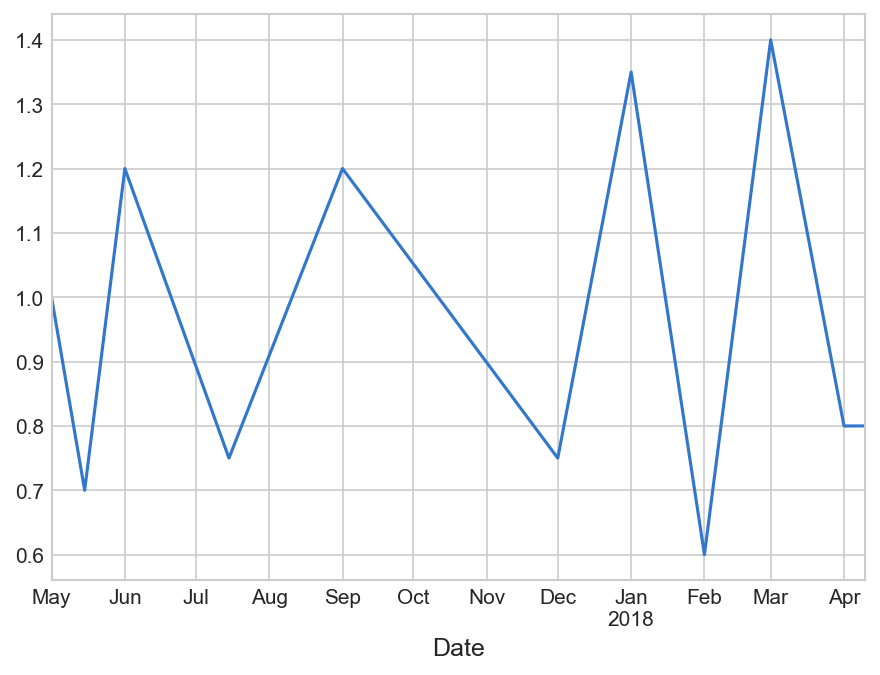}}
\caption{Expert correction term time series}
\label{expcorr}
\end{figure}

 Fig.~\ref{pred1} shows that in some time periods, predicted price is higher comparing to real price, in some time periods it is lower. 
 Fig.~\ref{targetpred} shows the ratio of real and predicted price. We can see that this ratio has periodic oscillations which describe the deviation of predictions from real values. 
It can be explained by existence of some factors which impact price but they are not included into the model (\ref{e1}). 
It can be the factors of complex behavior of investors. Suppose that some experienced expert understands such type of behavior. As a result, he or she can explain the deviation dynamics 
of regression model (\ref{e1}) forecasting from real values. Expert correction can be involved into the model by an additional term in the regression model  (\ref{e1}).
This term describes the dynamics of model deviation. To describe this term, the expert needs to define the local extremum on deviation time series which are pivot points for deviation trends. 
We suppose that the expert can define such points correctly relying on his or her personal experience. The time series for possible expert correction is shown on Fig. \ref{expcorr}.
\begin{figure}
\centerline{\includegraphics[width=1\textwidth]{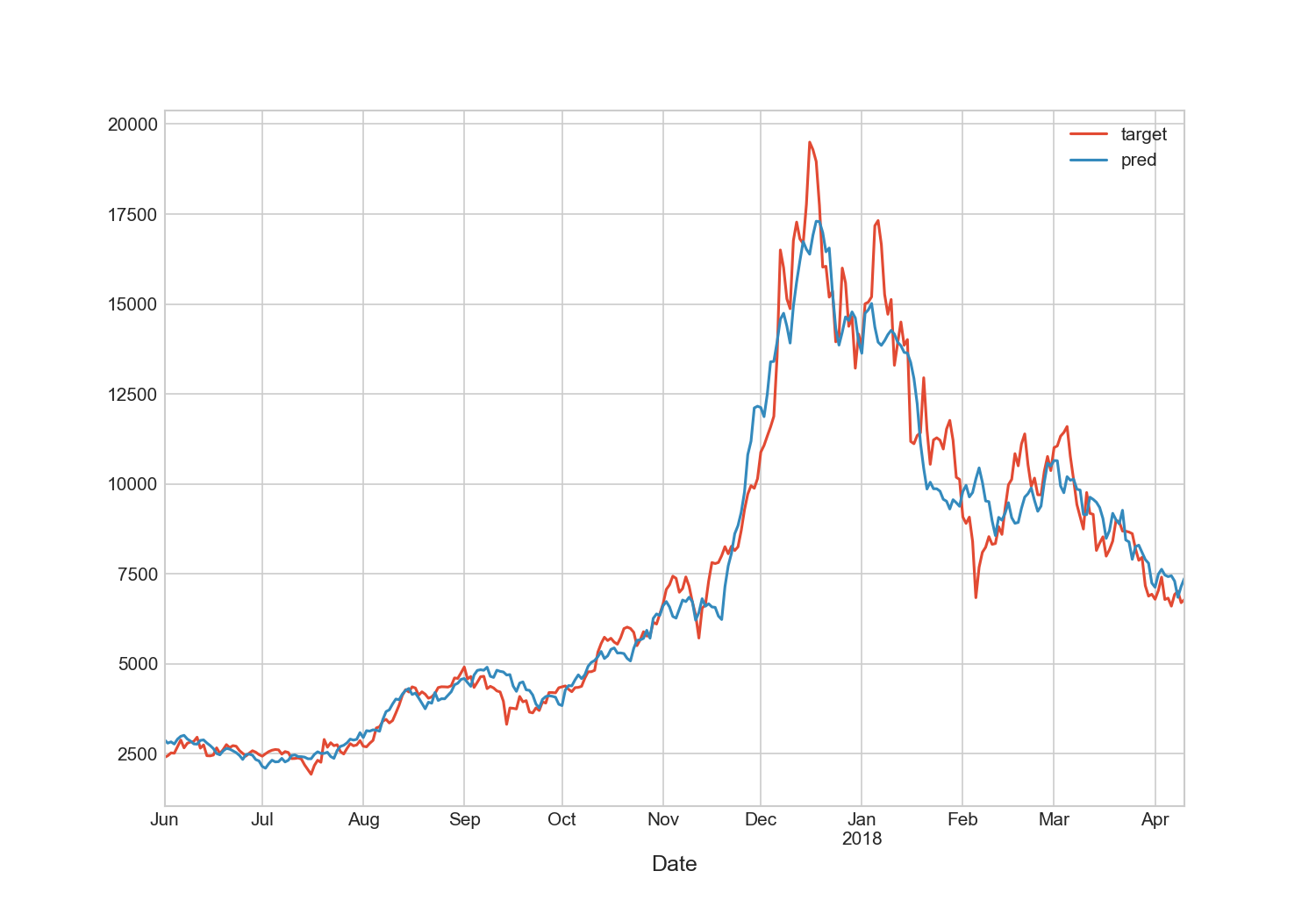}}
\caption{Bitcoin price dynamics \& prediction with expert correction}
\label{predcorr1}
\end{figure}
The results of the calculation of regression model  (\ref{e1}) with added such expert correction term are shown on the Fig.~\ref{predcorr1}.
For this case, we received the error value RMSE=856.4.  Fig.~\ref{featcoefcorr} shows the regression coefficient in case of expert correction term included into regression model.
Obtained results show that adding the expert term improves regression model performance. 
\begin{figure}
\centerline{\includegraphics[width=0.75\textwidth]{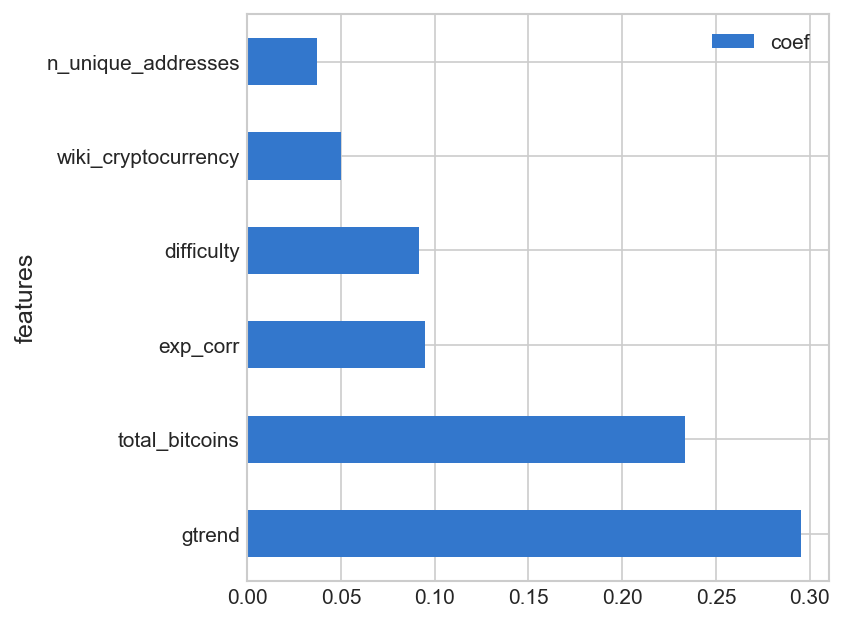}}
\caption{Feature coefficients for regression model with expert correction}
\label{featcoefcorr}
\end{figure}
\section{Bayesian regression model}
\begin{figure}
\centerline{\includegraphics[width=1\textwidth]{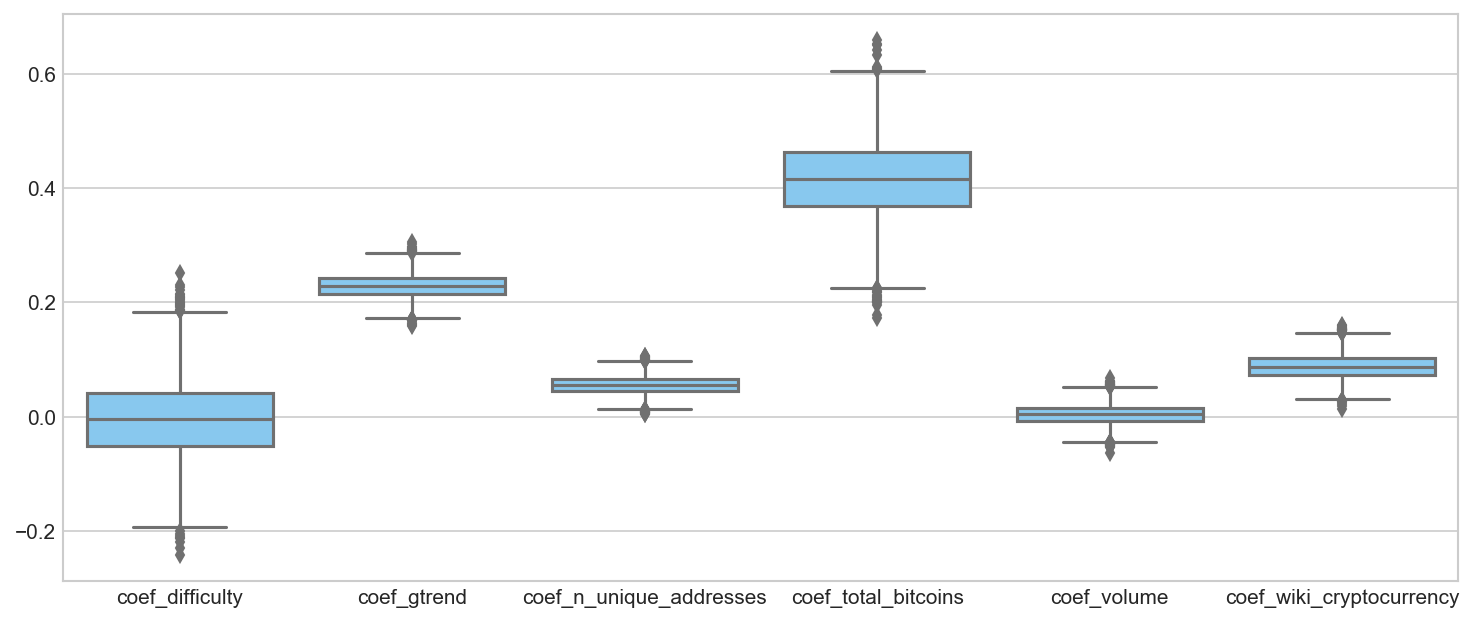}}
\caption{Boxplot for regression model coefficients}
\label{boxplot1}
\end{figure}

\begin{figure}
\centerline{\includegraphics[width=1\textwidth]{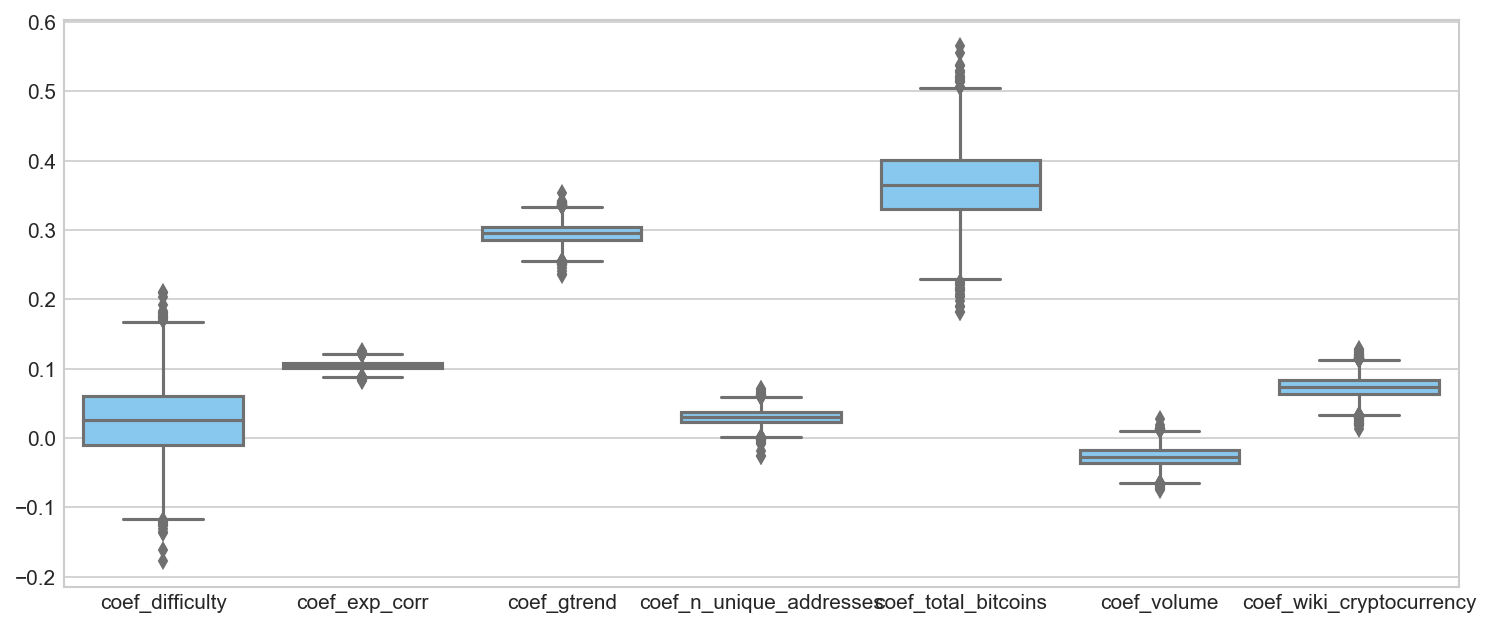}}
\caption{Boxplot for coefficients of regression model  with  expert correction term}
\label{boxplot1corr}
\end{figure}

For probabilistic approach, which makes it possible to get risk assessments, one can use Bayesian inference approach. 
Bayesian regression is a method which gives advantages when we need to take into account non Gaussian statistics for 
the variables under study (~\cite{kruschke2014doing}). 
For Bayesian modeling, we used Stan software with \textit{pystan} python package.
To do the model robust to outliers, we can consider Bitcoin price distributed with Student's t-distribution. 
Student's t-distribution is similar to normal distribution but it has heavier tails. 
Let us consider that the Bitcoin price is a stochastic variable which is distributed by Student's t-distribution
$$
price'_s \sim Student\_t(\mu,\sigma,\nu),
$$
where $\nu$ denote the degree of freedom, $\mu$ is a location parameter, $\sigma$ is a scale parameter. In Bayesian approach, we consider that 
$\mu=price'$, where $price'$ is defined by the regression model (\ref{e1}). 
Fig.~\ref{boxplot1} shows the box plots for received parameters of regression model.  Fig.~\ref{boxplot1corr} shows the box plots for the parameters in case of added expert correction term.
 For  scale parameter  $\sigma$, the value 0.14 was received in case of the absence of expert correction term and the value 0.1 was received in  case of added expert correction term. It shows that 
adding expert correction term into Bayesian regression model decreases the width of the probability distribution function for the target variable.

\section{Conclusion}
In our study, we considered the linear model for Bitcoin price which includes regression features based on Bitcoin currency statistics, 
mining processes, Google search trends, Wikipedia pages visits.
The pattern  of deviation of regression model prediction  from real prices is simpler comparing to price time series. So, we assume
that this pattern can be predicted by an experienced expert. In such a way, combining the regression model and 
expert correction, one can receive better results than with either regression model or expert opinion only.  
We assume that the expert can catch the complex time series patterns that relies on financial behavioral 
theories, economics and politics. These patterns cannot be caught from time series historical data since they exist for a short  time period. 
After these patterns are published or considered between investors, they soon become self-destructed. 
The received results show that the correct expert definition of time pivot points for the regression model deviation can essentially improve the prediction of Bitcoin price.  
In the proposed approach, the expert has to define time pivot points which describe the deviation of regression model based on the historical data comparing to real price time series.
With Bayesian inference, one can utilize the probabilistic approach using distributions with fat tails and take into account outliers in Bitcoin 
price time series. Having probability density distributions for price and model parameters, one can make risk assessment via calculating value at risk characteristics. 

\bibliographystyle{ieeetr}
\bibliography{ref1}
\end{document}